\documentclass[desactivate]{aa} 

\usepackage[varg]{txfonts}
\usepackage{orcidlink}
\usepackage{graphicx}   
\usepackage{hyperref}

\usepackage{natbib}
\bibpunct{(}{)}{;}{a}{}{,}

\newcommand{\response}[1]{{#1}}
\newcommand{\responsetwo}[1]{{#1}}

\defcitealias{Polak2023}{Paper I}

\begin{document} 

   \title{Massive star cluster formation}

   \subtitle{II. Runaway stars as fossils of subcluster mergers}

    \author{Brooke Polak
       \inst{1,2}\fnmsep\thanks{Fellow of the International Max Planck Research School for \newline Astronomy and Cosmic Physics at the University of Heidelberg}\orcidlink{0000-0001-5972-137X}
         \and
         Mordecai-Mark Mac Low\inst{2}\orcidlink{0000-0003-0064-4060}
         \and
        Ralf S. Klessen\inst{1,4}\orcidlink{0000-0002-0560-3172}
         \and
        Simon Portegies Zwart\inst{5}\orcidlink{0000-0001-5839-0302}
         \and
        Eric P. Andersson\inst{2}\orcidlink{0000-0003-3479-4606}
         \and
        Sabrina M. Appel \inst{6}\orcidlink{0000-0002-6593-3800}
         \and
        Claude Cournoyer-Cloutier\inst{7}\orcidlink{0000-0002-6116-1014}
         \and
        Simon C. O. Glover\inst{1}\orcidlink{0000-0001-6708-1317}
        \and
        Stephen L. W. McMillan\inst{3}\orcidlink{0000-0001-9104-9675}
       }

\institute{Universit\"{a}t Heidelberg, Zentrum f\"{u}r Astronomie, Institut f\"{u}r Theoretische Astrophysik, Heidelberg, Germany\\
           \email{bpolak@amnh.org}
        \and
         Department of Astrophysics, American Museum of Natural History, New York, NY, USA
         \and
         Department of Physics, Drexel University, Philadelphia, PA, USA
         \and
        Universit\"{a}t Heidelberg, Interdisziplin\"{a}res Zentrum f\"{u}r Wissenschaftliches Rechnen, Heidelberg, Germany
         \and
        Sterrewacht Leiden, Leiden University, Leiden, the Netherlands
         \and
        Department of Physics and Astronomy, Rutgers University, Piscataway, NJ, USA
        \and
         Department of Physics and Astronomy, McMaster University, Hamilton, ON, Canada
        }

   \date{Received 17 May 2024; accepted 5 August 2024}

 
  \abstract{  
   Two main mechanisms have classically been proposed for the formation of runaway stars. In the binary supernova scenario (BSS), a massive star in a binary explodes as a supernova, ejecting its companion. In the dynamical ejection scenario, a star is ejected during a strong dynamical encounter between multiple stars. We propose a third mechanism for the formation of runaway stars: the subcluster ejection scenario (SCES), where \response{a subset of stars from an infalling subcluster is ejected out of the cluster via a tidal interaction} with the contracting gravitational potential of the assembling cluster. We demonstrate the SCES in a star-by-star simulation of the formation of a young massive cluster from a $10^6\rm\, M_\odot$ gas cloud using the \textsc{torch} framework. 
   This star cluster forms hierarchically through a sequence of subcluster mergers determined by the initial turbulent, spherical conditions of the gas. We find that these mergers drive the formation of runaway stars in our model. Late-forming subclusters fall into the central potential, where they are \response{tidally disrupted}, forming \response{tidal tails} of runaway stars that are distributed highly anisotropically. 
   Runaways formed in the same SCES have similar ages, velocities, and ejection directions.
   Surveying observations, we identify several SCES candidate groups with anisotropic ejection directions.
   The SCES is capable of producing runaway binaries: two wide dynamical binaries in infalling subclusters were tightened through ejection. This allows for another velocity kick via subsequent via a subsequent BSS ejection. 
   An SCES-BSS ejection is a possible avenue for the creation of hypervelocity stars unbound to the Galaxy.
   The SCES occurs when subcluster formation is resolved. We expect nonspherical initial gas distributions to increase the number of calculated runaway stars, bringing it closer to observed values. The observation of groups of runaway stars formed via the SCES can thus reveal the assembly history of their natal clusters.     
    }


   \maketitle
%
\section{Introduction}

Young stars displaced from their birthplace in the star-forming spiral arms of the Galaxy and moving away from the Galactic disk were first observed by \citet{Blaauw1954ApJ...119..625B}. Since then, many more stars moving rapidly away from their formation sites have been observed throughout the Galaxy. These stars are typically classified as runaways when their velocity relative to their associated nebula or cluster is $\ge30\mbox{ km s}^{-1}$ \citep{Gies1986ApJS...61..419G}.

Despite their ubiquity, the ejection mechanism remains unknown for most runaway stars. Currently, there are two popular proposed mechanisms for producing runaway stars: the binary supernova scenario \citep[BSS; sometimes referred to as Blaauw kicks;][]{Blaauw1961BAN....15..265B} and the dynamical ejection scenario \citep[DES;][]{Poveda1967BOTT....4...86P,Hoogerwerf2000ApJ...544L.133H, Fujii2011Sci...334.1380F}. According to the BSS, when two massive stars are in a binary system and one explodes as a supernova (SN), 
  \response{loss of the ejecta from the system}
reduces the gravity on the companion, which begins moving through space at a velocity comparable to its orbital velocity. According to the DES, runaway stars are ejected during a strong dynamical encounter involving at least one binary, in which orbital binding energy is converted to kinetic energy. In some cases, a runaway system is formed moving in the opposite direction of the system that ejected it \citep[in the center-of-mass frame of the encounter;][]{Poveda1967BOTT....4...86P, Fujii2011Sci...334.1380F}. 
\response{BSS and DES runaways can be distinguished by their rotational and linear velocities: the BSS produces slow-moving, rapidly rotating stars, while the DES produces fast-moving, slowly rotating stars \citep{Sana2022A&A...668L...5S}.}

In this work we propose a third mechanism for producing runaway stars: the subcluster ejection scenario (SCES), in which \response{part of} a subcluster is ejected from the cluster after \response{it is tidally disrupted by} the contracting potential of the assembling cluster. Star clusters form hierarchically, with giant molecular clouds (GMCs) fragmenting into dense clumps that form subclusters of stars. These subclusters merge, forming a single central cluster. The combined feedback eventually blows away the unused gas, leaving a gas-free star cluster
\citep[see, e.g.,][]{Rahner2017MNRAS.470.4453R,Grudic2018MNRAS.481..688G,Rahner2019MNRAS.483.2547R,wall2019,2020Wall,2021Cournoyer-Cloutier,Lewis2023ApJ...944..211L,Wilhelm2023MNRAS.520.5331W,Cournoyer-Cloutier2023MNRAS.521.1338C}. \responsetwo{\citet{Lucas2018MNRAS.474.3582L} shows through $N$-body simulations that merging groups of stars can produce a tidal tail of unbound stars. The SCES describes when this mechanism occurs between young subclusters during initial cluster assembly.} To thoroughly investigate the origin of SCES runaway stars, the entire complex dynamical history of young star clusters must be modeled consistently. 
This necessitates modeling the entire star cluster formation process from the birth of stars within subclusters. 

We present a star-by-star simulation of cluster formation performed with the \textsc{torch}\footnote{Version used for this work: \href{https://bitbucket.org/torch-sf/torch/commits/tag/massive-cluster-1.0}{https://bitbucket.org/torch-sf/torch/commits/tag/massive-cluster-1.0}} framework \citep{wall2019,2020Wall}. \textsc{torch} follows gas dynamics, the $N$-body dynamics of stars, sub-grid star formation via sink particles, stellar evolution, and stellar feedback in the form of winds, radiation, and SNe. With this simulation, we can determine the origin of runaway stars formed self-consistently in the cluster environment. 

In this paper we analyze the runaway stars formed in the M6 star cluster presented in \citet[hereafter \citetalias{Polak2023}]{Polak2023}, which forms from a molecular cloud with an initial mass of $M_{\rm cloud} = 10^6\rm\, M_\odot$ and an initial radius of $R_{\rm cloud} =11.7\rm\,pc$. We briefly describe our model and the initial conditions in Sect.~\ref{section:methods}. In Sect.~\ref{section:results} we present the properties of the runaway stars formed in M6, as well as our evidence that the SCES mechanism produces runaway stars. We discuss the implications of our results in Sect.~\ref{section:discussion} and present some observational candidates for runaways formed via the SCES. We conclude in Sect.~\ref{section:conclusions}. 

\section{Methods}
\label{section:methods}

\begin{table}
\caption{Simulation parameters.}   
\label{table:params}   
\centering   
\begin{tabular}{l|cc}          
\hline\hline                        
Parameter & Value & Units \\   
\hline 
    $M_\mathrm{cloud}$ & $10^6$ & M$_\odot$ \\
    $\rho_{c}$ & 280 & M$_\odot$ pc$^{-3}$\\
    $\Bar{\rho}$ & 150 & M$_\odot$ pc$^{-3}$\\    
    $\Sigma\ $ & 2325 & M$_\odot$ pc$^{-2}$ \\
    $B_{0,z}$ & 18.5 & $\mu$G \\
    $\lambda_{\rm J}$ & 1.0 & pc\\
    $t_\mathrm{ff}$ & 0.67 & Myr\\ 
    $R_\mathrm{cloud}$ & 11.7 & pc\\
    $R_\mathrm{box}$ & 20.0 & pc\\ 
    $\alpha_\mathrm{v}=E_{\rm kin}/|E_{\rm pot}|$ & 0.15 & -\\
    $\Bar{v}_{\rm gas}$ & 8.15 & km~s$^{-1}$ \\
    $\sigma_{v\rm\,, gas}$ & 3.25 & km~s$^{-1}$ \\
    $\Delta$x$_\mathrm{min}$ & 0.3125 & pc\\
    $\Delta$x$_\mathrm{max}$ & 1.25 & pc\\
    $r_\mathrm{sink}$ & 0.78125 & pc\\
    $\rho_\mathrm{sink}$ & $8\times 10^{-21}$ & g cm$^{-3}$\\ 
    $M_\mathrm{sink}$ & 246 & M$_\odot$\\ 
    $M_\mathrm{feedback}$ & 20 & M$_\odot$\\ 
    $M_\mathrm{n-body}$ & 4 & M$_\odot$\\ 
    $M_\mathrm{IMF}$ & 0.08--100 & M$_\odot$\\ 
\hline  
\end{tabular}
\tablefoot{Rows: initial cloud mass, central density, average density, surface density, peak initial vertical magnetic field (see Eq.~\ref{eq:mag_field}), initial central Jean's length, initial free-fall time, cloud radius, half-width of box, virial parameter, average velocity and velocity dispersion of the gas within $R_{\rm cloud}$, minimum cell width, maximum cell width, sink radius, sink threshold density, approximate initial sink mass, minimum feedback star mass, agglomeration mass of low-mass stars, mass sampling range of Kroupa IMF.}
\end{table}

We used \textsc{torch} \citep{wall2019, 2020Wall} to model the formation of a star cluster from a turbulent spherical cloud of gas. The \textsc{torch} star cluster formation framework is built with the Astrophysical MUltpurpose Software Environment \citep[AMUSE;][]{PORTEGIESZWART2009369amuse1,Portegies2013CoPhC.184..456Pamuse2, amuse, amusebook} framework, linking several physics codes addressing hydrodynamics, stellar evolution, stellar dynamics, star formation via sink particles, and stellar feedback in the form of winds, radiation, and SNe. In this section we provide a brief overview of the numerical methods used for these simulations. A more detailed description of the applied methods is provided in \citetalias{Polak2023}.

The adaptive mesh refinement (AMR) magnetohydrodynamics code \textsc{flash} \citep{flash} handles the gas dynamics. The gas dynamics are evolved with an HLLD Riemann solver \citep{Miyoshi2005JCoPh.208..315M} and third-order piecewise parabolic method reconstruction \citep{Colella1984JCoPh..54..174C}. The self-gravity of the gas is evolved with a multi-grid Poisson solver \citep{Ricker2008ApJS..176..293R}, and the gravitational interaction between the stars and gas is modeled using a leapfrog gravity bridge scheme \citep{Fujii2007PASJ...59.1095F,wall2019}.

\textsc{torch} uses a sub-grid star formation model via sink particles in \textsc{flash}. Sink particles form when the local gas density exceeds a threshold value $\rho_{\rm sink}$ based on the Jeans density at the maximum refinement level, and when several other criteria are met \citep[see][]{Federrath_2010}. When a sink particle forms, the gas within the sink accretion radius $r_{\rm sink}$ and above the threshold density is added to the sink's mass. For each sink, the Kroupa initial mass function \citep[IMF;][]{2002Sci...295...82Kroupa} is randomly sampled to create a list of stellar masses for the sink to form sequentially. From this list, the sink forms star particles until it runs out of mass. The sink continues to form stars whenever it accretes enough mass to form the next star on the list. When stars form, they are placed randomly in a uniform spherical distribution within the sink accretion radius and given a velocity equal to the sink's velocity plus an additional velocity component with the direction sampled isotropically and the magnitude sampled from a Gaussian distribution with a standard deviation equal to the local sound speed. The local sound speed is taken to be the average sound speed of gas with temperature $T \le 100\rm\, K$ in cells  $\le 2r_{\rm sink}$ from the sink. If no gas fulfills this criterion within the region, the sink is blocked from forming stars. Sink momentum is set by adding the current sink momentum to the momentum of the gas it accretes.

Stellar feedback in the form of stellar winds, radiation, and SNe, are also implemented as additional physics in \textsc{flash} \citep{2020Wall}. Radiative transport is modeled in Flash using the ray-tracing routine Fervent \citep{FERVENT10.1093/mnras/stv1906}. Stars are evolved from the zero-age main sequence until their death via \textsc{SeBa} \citep{seba}, which informs the stellar feedback properties in \textsc{flash}. \textsc{petar} \citep{petar} evolves the N-body dynamics of the star particles. 

To reduce the computational expense caused by the large number of stars formed in a $10^6\rm~M_\odot$ cloud, we applied three modifications to the standard \textsc{torch} model: 
\begin{enumerate}
    \item Star particles forming below $M_{\rm agg}=4\rm\, M_\odot$ are agglomerated until their summed mass is above $M_{\rm agg}$ and then are assigned to a single super-star particle. This reduces the number of stars from $\sim 10^6$ to $\sim 10^5$ to make the N-body calculations feasible. 
    \item The mass-loading of stellar winds is artificially raised to limit the temperature of wind-blown bubbles to $T_{\rm w}=3\times 10^5\rm\, K$ thereby alleviating the timestep constraint set by the Courant condition. Although this causes wind bubbles to be smaller and cooler, the primary effect of winds is to clear out dense gas, allowing ionizing radiation to escape and form \ion{H}{II} regions. The clearing of dense gas still occurs with the mass-loaded winds. At the densities in our simulation, hot stellar winds rapidly cool regardless of mass-loading.
    \item Ray-tracing on the AMR grid quickly becomes expensive for a large number of sources. To reduce the cost of the ray-tracing calculations, we only include feedback (radiation, winds, and SNe) from stars with $M_{\rm feedback} \ge 20\rm\, M_\odot$. Most of the mechanical wind energy and ionizing radiation in a star cluster comes from massive stars, so with this limit $<20\%$ of the total feedback energy is lost (see \citetalias{Polak2023}). SNe of stars below this mass limit take $\ge 10\rm\, Myr,$ which is much longer than the scope of these simulations.
\end{enumerate}
A detailed discussion of the effect of these modifications is provided in \citetalias{Polak2023}.

We used the M6 simulation introduced in \citetalias{Polak2023} of initial mass $M_{\rm cloud} = 10^6 \rm\, M_\odot$, with properties listed in Table \ref{table:params}. The initial density profile is a Gaussian \citep{Bate1995MNRAS.277..362B,Goodwin2004A&A...414..633G} with $\rho_{\rm edge}/\rho_{\rm c}=1/3$. We used outflow boundary conditions that allow inflow from ghost zones. The turbulence was set by imposing a \cite{Kolmogorov1941DoSSR..30..301K} velocity spectrum on the gas. The initial magnetic field, $\vec{B} = B_z \hat{z}$, is uniform in $z$ and decreases radially with the mid-plane density $\rho(x,y,z=0)$ in the $x$-$y$ plane:
\begin{equation}
\label{eq:mag_field}
    B_z(x,y) = B_{0,z} \exp\left[ - (x^2+y^2) \ln(3) / R_\mathrm{cloud}^2 \right]
,\end{equation}
where $B_{0,z} = 18.5\,\mathrm{\mu G}$.

\section{Results}
\label{section:results}

In our analysis, we made an initial selection for runaway stars by filtering for unbound stars,\footnote{Canonically, runaway stars are defined as stars leaving their birthplace at velocities $\ge 30\rm\, km~s^{-1}$. Unbound stars with lower velocities are typically referred to as walkaways. All unbound stars we consider in this work have velocities $\ge 30\rm\, km~s^{-1}$ criterion.} with $E_{\rm total,i}=E_{k,i}+U_{\star,ij}+U_{\rm gas}
>0,$ where $E_{k,i}$ is the kinetic energy, $U_{\star,ij}$ is the gravitational potential of other star particles, and $U_{\rm gas}$ is the gravitational potential of the gas. Our analysis only covers the early runaways ejected during the hierarchical assembly of the cluster. 
The free-fall time of the M6 cloud is $t_{\rm ff}=0.67$ Myr, and the final simulation time is $1.36t_{\rm ff}=0.91$ Myr. The cluster is fully assembled by the end of the simulation, and the SCES only occurs during active subcluster formation. This run time is therefore sufficient for analysis of the SCES.
At the final time, the fraction of runaway stars is $0.51\%$. We cannot yet determine the final runaway star fraction due to the short run time of the simulation. Simulating the total runaway fraction requires integration times of $\gtrsim 2\rm~Myr$, which is outside the scope of this paper introducing a production mechanism for early-forming runaway stars in young clusters.

\begin{figure}[t]
\centering
    \includegraphics[width=1\hsize]{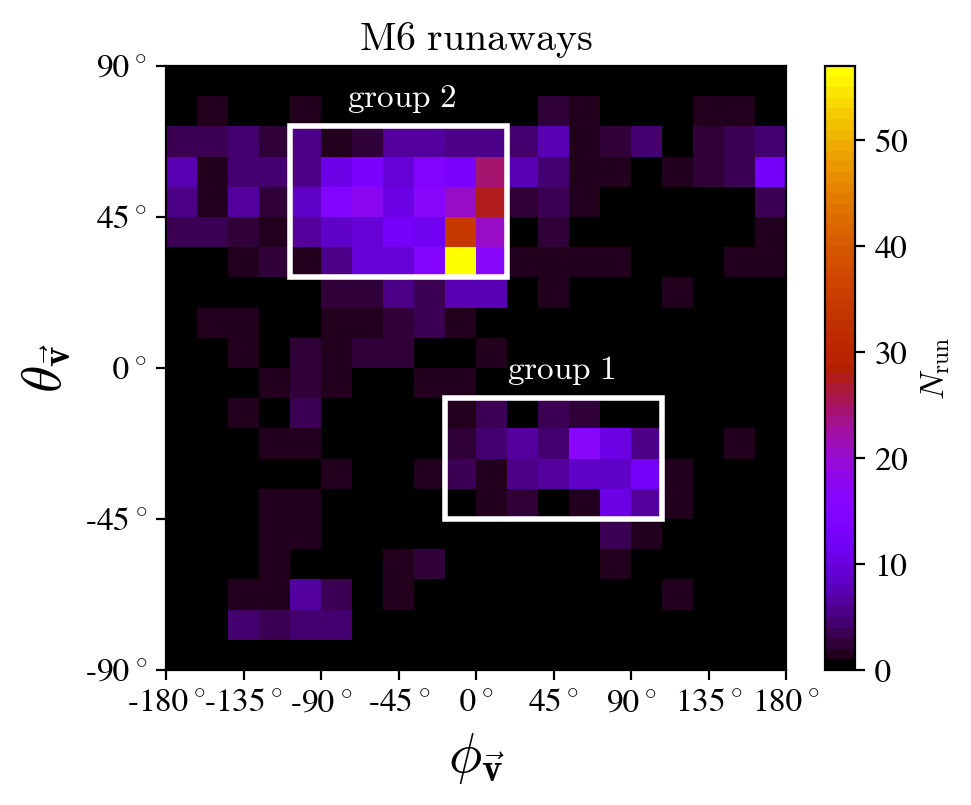}
    \caption{Ejection velocity angular directions of M6 runaway stars with ${\theta}_{\vec{v}}$ and ${\phi}_{\vec{v}}$ binned in $9^\circ$ and $18^\circ$ angle bins, respectively. There are two distinctly peaked angular regions, which are labeled as group 1 and group 2. The rectangles indicate the angular bins of runaways we selected for trajectory analysis.}
    \label{fig:M6_run_hist}
\end{figure}

\begin{figure}[t]
\centering
    \includegraphics[width=1\hsize]{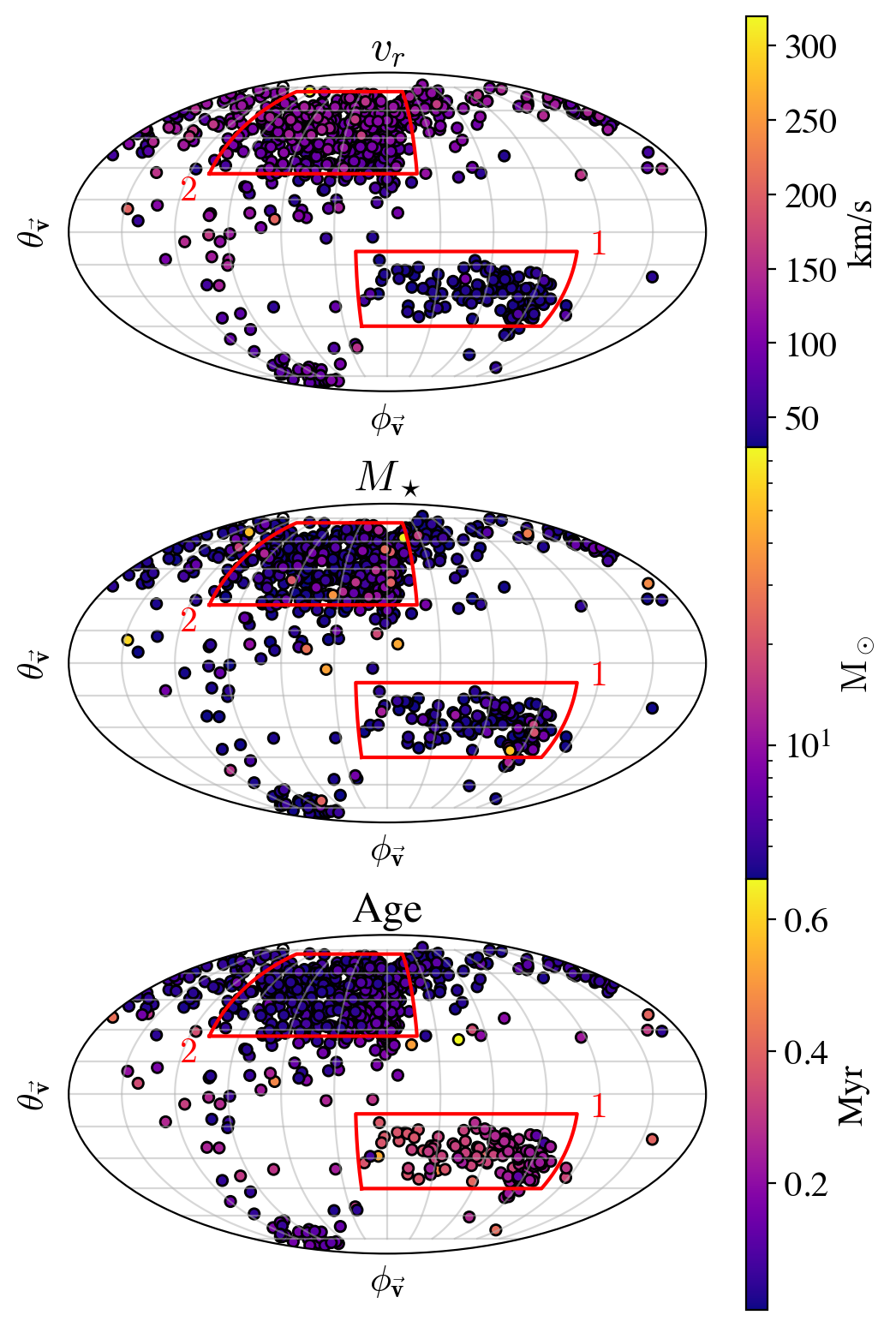}
    \caption{Mollweide map projection of the ejection directions of the runaway stars in M6. The colors correspond to the radial velocity from the cluster center of mass (top), the star mass (middle), and the star age (bottom). The highest values are plotted over lower values for visibility.}
    \label{fig:M6_run_map}
\end{figure}

\begin{figure}[t]
\centering
    \includegraphics[width=1\hsize]{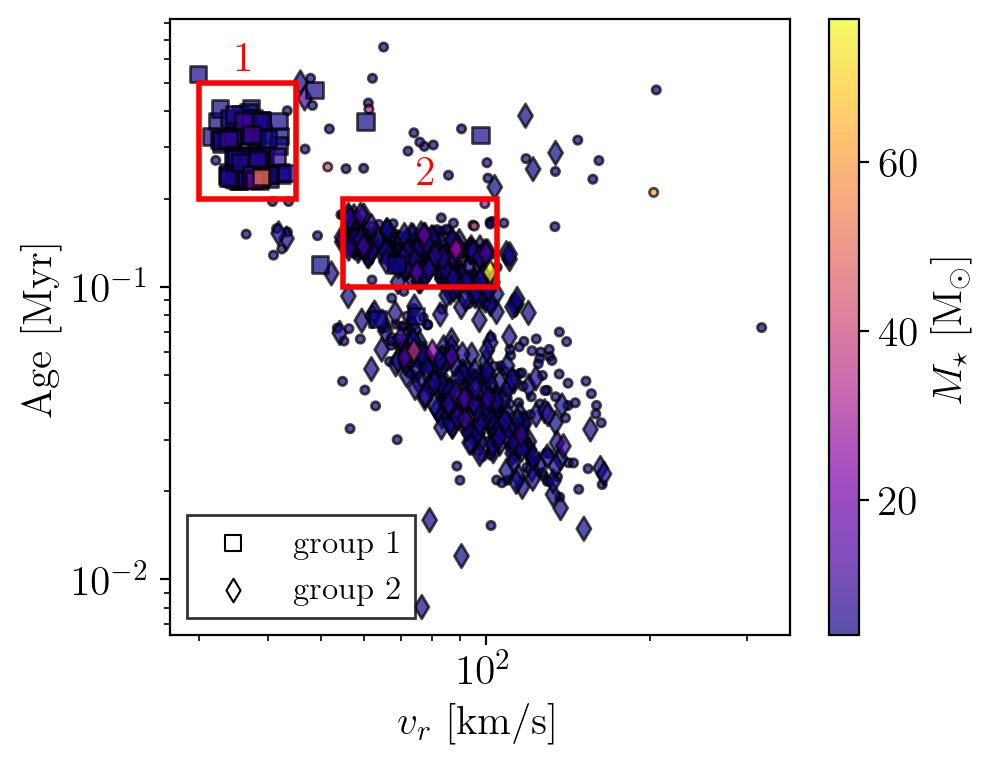}
    \caption{Scatter plot of the runaway stars, with the radial velocity and age colored by mass. Squares are group 1, diamonds are group 2, and dots are the rest of the runaways. These groups are the runaways selected by ejection direction only. The two groups are distinctly separate in both age and velocity. Group 2 contains two subgroups ejected in the same direction but at slightly different times. For the purpose of demonstrating the SCES mechanism, we focus on the later forming subgroup in group 2. The red rectangles indicate the additional filtering of the runaway groups by age and velocity.}
    \label{fig:M6_age_vel}
\end{figure}

\begin{figure}[t]
\centering
    \includegraphics[width=1\hsize]{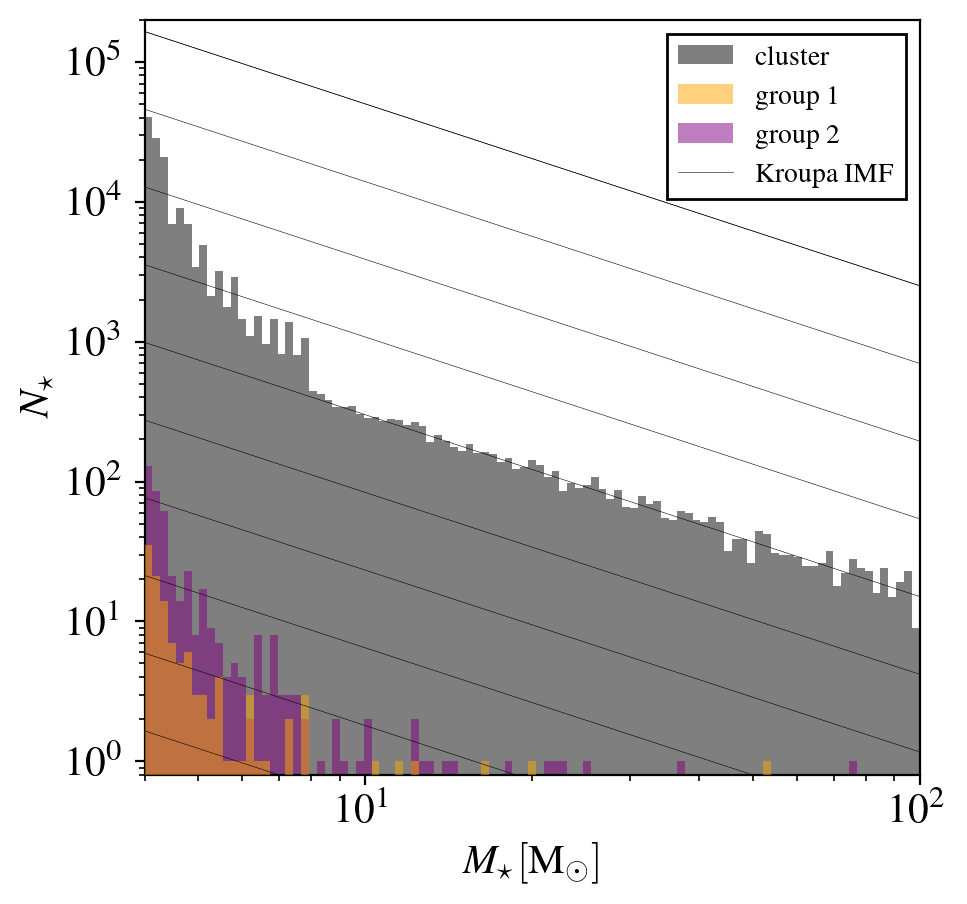}
    \caption{Particle mass distributions of the entire star cluster (gray) and the runaway groups along with the Kroupa IMF power law. Note that at the low-mass end ($\le8\rm M_\odot$), the particle masses are skewed higher than the sampled IMF due to the agglomeration of stars below $<4\rm M_\odot$. }
    \label{fig:imf}
\end{figure}

\begin{figure}[t]
\centering
    \includegraphics[width=1\hsize]{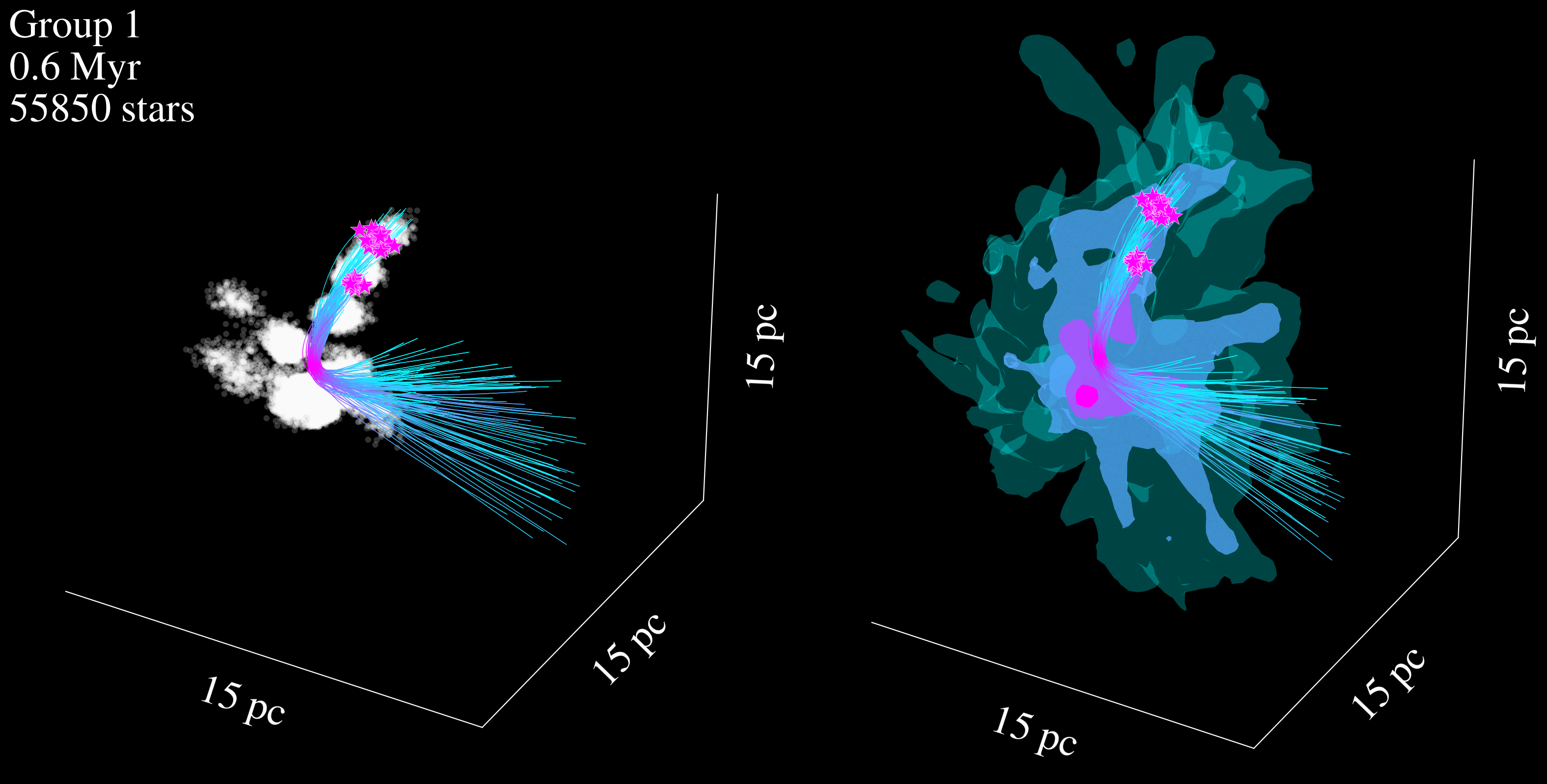}
    \caption{Movie of the formation and ejection of runaway star group 1. Left: Stars as they are forming in the cluster.\ Right: Gravitational potential of both the stars and gas. The trajectories of the runaway stars are shown as lines. Once each runaway star forms, it appears with a star marker. This plot shows the same variables as shown in Figs.~\ref{fig:M6_group1_traj} and~\ref{fig:M6_traj_potential}.  }
    \label{fig:movie}
\end{figure}

\begin{figure*}[t!]
\centering
    \includegraphics[width=0.8\textwidth]{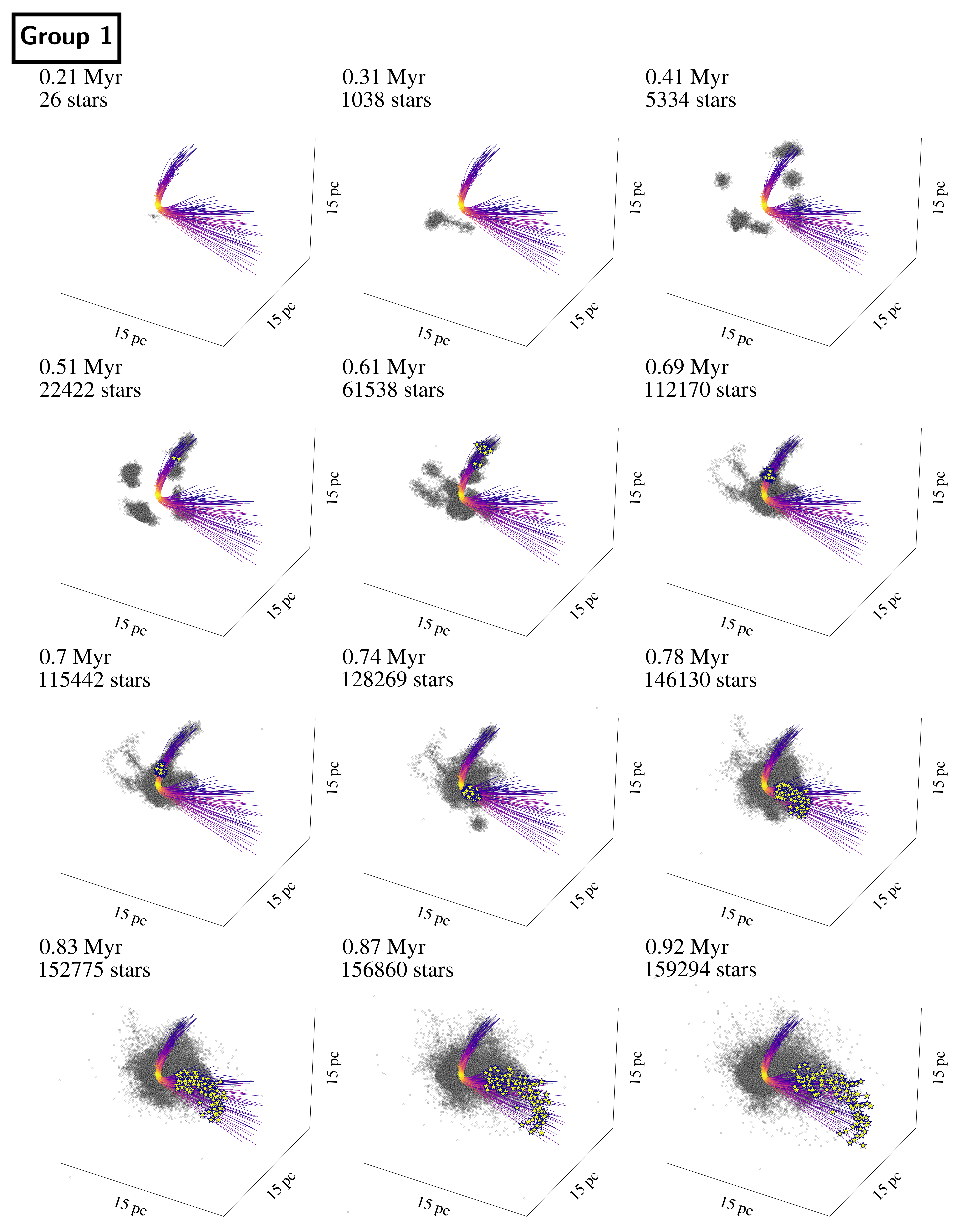}
    \caption{Trajectories of runaway star group 1 originating from a subcluster merging into the central cluster, with color indicating the stellar velocity (light$=v_{\rm max}$, dark$=v_{\rm min}$). The dots are star particles at the given time, and the star markers indicate the runaway stars in the group as they form. The total number of stars in the cluster is listed in each panel.}
    \label{fig:M6_group1_traj}
\end{figure*}

\begin{figure*}[t!]
\centering
    \includegraphics[width=0.8\textwidth]{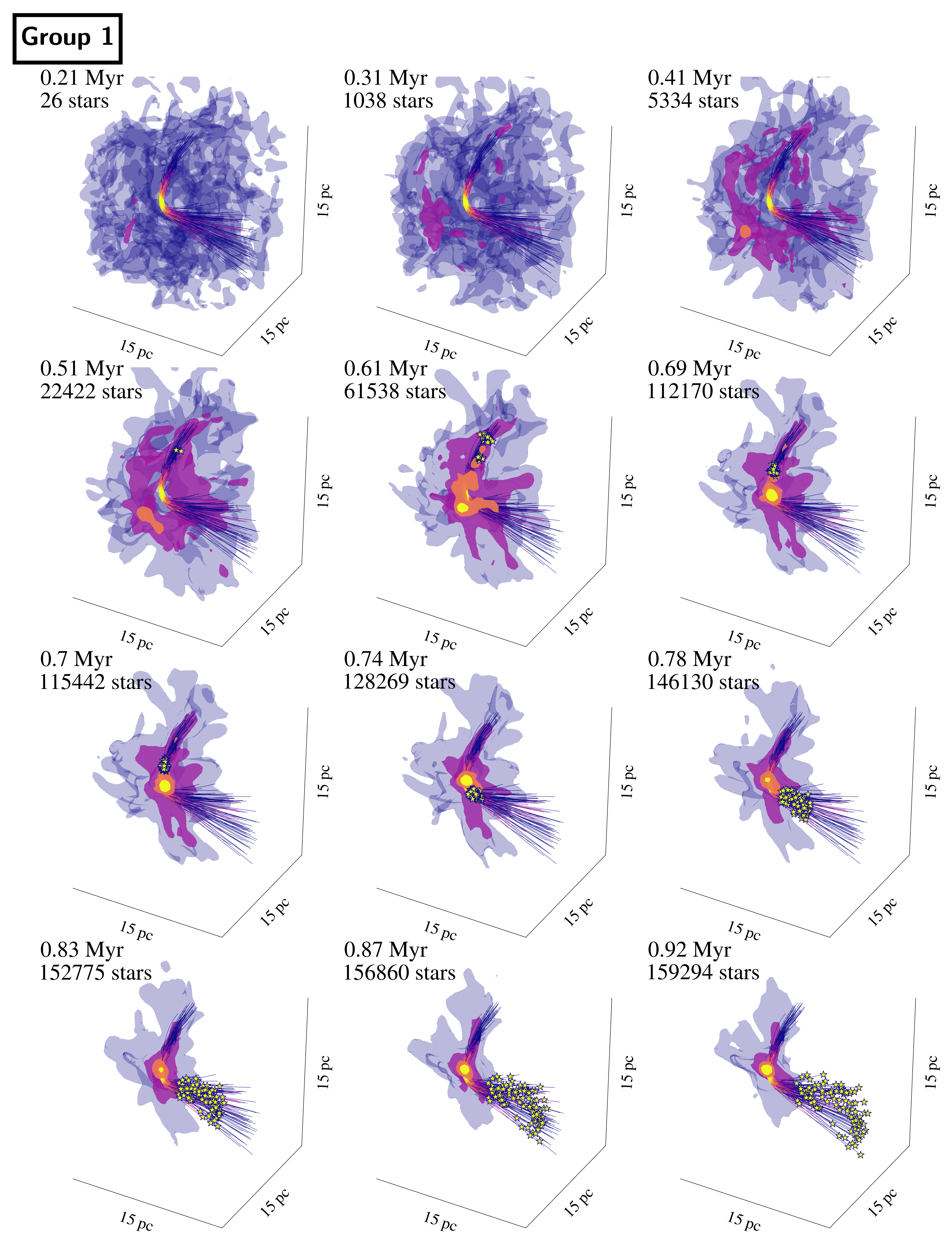}
    \caption{Trajectories of runaway star group 1 originating from a subcluster merging into the central cluster, with color indicating the stellar velocity (light$=v_{\rm max}$, dark$=v_{\rm min}$). The isosurfaces show four values of the gravitational potential of both the stars and gas, $|U_g|=10^{47,48,49,50}\rm erg$, with dark to light from lowest to highest. As the runaway stars form, they are plotted with star markers. The total number of stars in the cluster is listed in each panel.}
    \label{fig:M6_traj_potential}
\end{figure*}

\begin{figure*}[t!]
\centering
    \includegraphics[width=0.925\textwidth]{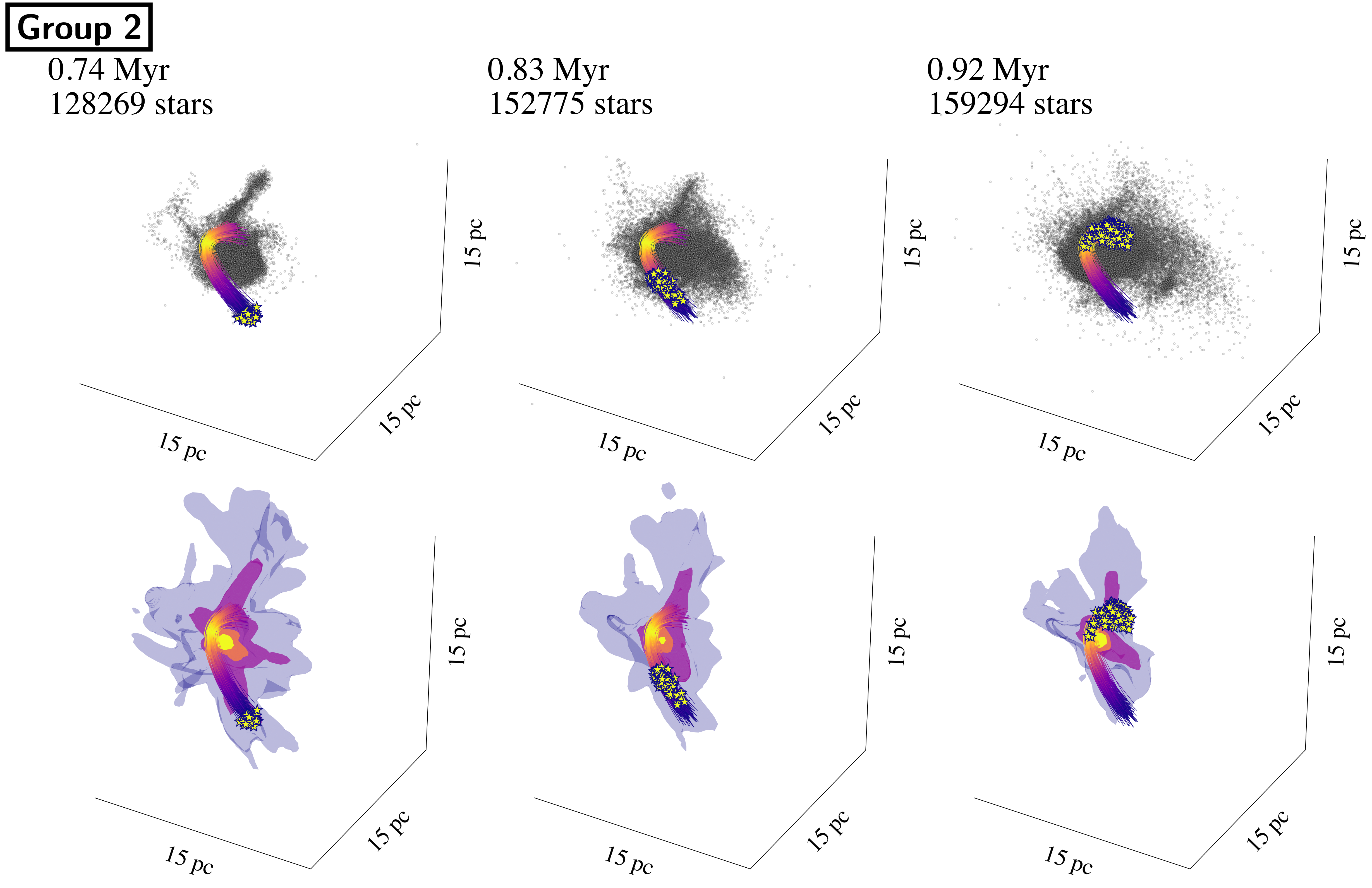}
    \caption{Trajectories of runaway star group 2 originating from a subcluster merging into the central cluster, with color indicating the stellar velocity (light$=v_{\rm max}$, dark$=v_{\rm min}$). The dots in the top row (analogous to Fig.~\ref{fig:M6_group1_traj}) represent star particles, and the star markers indicate the runaway stars in the group as they form. The bottom row (analogous to Fig.~\ref{fig:M6_traj_potential}) shows the same runaway star trajectories, with the potential of gas and stars plotted as isosurfaces. The isosurfaces show four values of the gravitational potential of both the stars and gas, $|U_g|=10^{47,48,49,50}\rm erg,$ with dark to light from lowest to highest.}
    \label{fig:M6_group2_traj}
\end{figure*}

Figure \ref{fig:M6_run_hist} shows a histogram of the runaway star count binned by the angular direction of the runaway velocities in the center-of-mass frame. There are two distinct groupings of runaway stars ejected in the same directions, and the overall directional distribution of runaways is highly anisotropic. \response{This is caused by the tidal disruption of two infalling subclusters by the cluster center of mass, which then form two groups of runaway stars as tidal tails.} White boxes in Fig. \ref{fig:M6_run_hist} indicate the selection regions used to investigate the origin of these two groups of runaways. These regions were chosen by eye with the intent of performing a broad initial analysis. 

The properties of the runaway stars are depicted in Fig. \ref{fig:M6_run_map}, which shows a Mollweide spherical projection of the runaway stars colored according to radial velocity from the center-of-mass, mass, and age. The group 2 population is moving faster than group 1. The average velocity and standard deviation for group 1 and 2 are $\Bar{v_1}=38.6\rm~km~s^{-1},\ \sigma_1=8.2\rm~km~s^{-1},\ \Bar{v_2}=87.1\rm~km~s^{-1},\ \sigma_2=20.7\rm~km~s^{-1}$. Both groups are composed mostly of low-mass stars with a few high-mass stars scattered within. The stars in a given group were formed at similar ages.

Figure \ref{fig:M6_age_vel} shows the age and velocity of the runaways in a scatter plot colored according to stellar mass, with the selected angle bins in Fig. \ref{fig:M6_run_hist} marked by symbol type. The runaways are distinctly clustered in both radial velocity and stellar age. Group 1 is distinguished from the rest of the runaways in the upper left; stars in group 1 are older and have lower velocities ($v_r\approx30\rm~km~s^{-1}$). Group 2, however, consists of two populations of runaways seen as an age gap in the diamond points in Fig.~\ref{fig:M6_age_vel}. This indicates stars from two subclusters or two generations of stars within one subcluster were ejected in the same direction. The mass distributions of both groups are the same, though, with both mostly containing low mass stars and a few stars above $10\rm\, M_\odot$. Figure~\ref{fig:imf} shows the Kroupa IMF power law with $\alpha=-2.3$ and the histograms of the stellar masses in each runaway group as well as the entire cluster. The overpopulation of low mass bins ($<8\rm~M_\odot$) is a result of the mass agglomeration of stars below $<4\rm\, M_\odot$. This suggests that each group consists of a random sampling of the IMF, which makes sense for a subcluster. Both runaway groups roughly resemble the IMF indicating that the SCES mechanism is unbiased toward stellar mass. We propose that this is because the stars in the subcluster closest to the potential are \response{tidally} captured while the outer stars are ejected, and the spatial distribution of stellar masses within a subcluster is random by construction. Most of the runaway stars are agglomerated, which is expected from a sampling of the broader stellar population. This is acceptable as there is no apparent mass preference or cutoff for ejection via the SCES.


\begin{table}
\caption{Runaway group filters.}   
\label{table:run_select}   
\centering   
\begin{tabular}{l|cc}          
\hline\hline                        
Parameter & Group 1 & Group 2 \\   
\hline                                   
   $\phi_{\vec{v}}$ & $[-18^\circ , 108^\circ]$ & $[-108^\circ , 18^\circ]$ \\
   $\theta_{\vec{v}}$ & $[-45^\circ , -9^\circ]$ & $[27^\circ , 72^\circ]$ \\
   $v_r$ & $[30,45]$ km/s & $[55,105]$ km/s\\
   $\tau$ & $[0.2,0.5]$ Myr & $[0.1,0.2]$ Myr \\
\hline  
\end{tabular}
\tablefoot{Selection ranges for runaway stars in groups 1 and 2. Rows: $\phi$ and $\theta$ directions of velocity, radial velocity from center of mass, and stellar age.}
\end{table}

We applied a second filter to select the runaways belonging to the two groups using stellar age and radial velocity indicated in Fig. \ref{fig:M6_age_vel}. We selected the younger of the two populations in group 2 to focus on a specific subcluster. The three most important characteristics for identifying groups of runaways are ejection direction, velocity, and stellar age. The complete list of filters for groups 1 and 2 is listed in Table \ref{table:run_select}. We find that selecting groups based on direction, age, and radial velocity provides a sufficient selection of runaways from a particular subcluster merger. 

Using these age, velocity, and direction filters, we traced the stars in the two groups back through the evolution and formation of the star cluster to determine their ejection mechanism. Figure~\ref{fig:movie} is a still from a movie (available in the online version of the paper) showing the formation and ejection of runaway group 1 as well as the formation of the star cluster. In addition to the angular selection outlined in Table~\ref{table:run_select}, we also imposed an age restriction to select stars that formed at the same time. In this way, we selected only one of the two age groups in group 2. 

Figures~\ref{fig:movie}-\ref{fig:M6_group2_traj} reveal \response{tidal interactions of the subcluster} with the main cluster as the physical mechanism for ejecting these two groups, explaining their shared ejection direction, age, and velocity as well as their uniformly sampled IMF mass composition. Group 1 and 2 stars form from distinct subclusters. The subcluster merges with the assembled central massive cluster \response{becoming tidally disrupted. Some stars are tidally captured while others are slingshot around the central potential and escape as tidal tails.} 

The timing of the contraction and expansion of the central cluster potential is crucial for the SCES. In a static potential, the stars falling in could not escape the cluster due to conservation of energy. In the SCES, a subcluster approaches the central potential as the potential contracts due to the assembly of other subclusters and gas infall. After the \response{tidal interaction}, as the SCES stars are leaving the central cluster, the potential well expands as star formation from the gas proceeds and the cluster virializes. The timing of the subcluster formation and cluster potential evolution determines whether the SCES stars become unbound.

Group 2 is more populous than group 1. This is not because subcluster 2 formed more stars, but because the potential driving the \response{tidal force} on group 2 is stronger than when group 1 fell in, resulting in more stars being ejected. From Fig.~\ref{fig:M6_group1_traj} we see many stars that do not become runaways forming in the vicinity of group 1 stars, whereas the majority of stars forming near group 2 stars in Fig.~\ref{fig:M6_group2_traj} become runaways.

We find that SCES runaways from later interactions with a deeper potential well result in a higher fraction of stars ejected within a subcluster. The \response{tidal interaction} is highlighted further by the figures showing trajectories with the evolving potential. At the onset of infall for group 1, the potential is significantly shallower than when group 2 forms. The peak of the potential has just formed and is still migrating when group 1 forms, but it is at its final destination when group 2 forms. However, the potential reaches its peak by the time of slingshot (point of maximum stellar velocity) for both groups. The greater integrated acceleration results in more stars in subcluster 2 being ejected than in subcluster 1. This correlation allows us to probe the assembly history of a star cluster from the ejection directions, ages, and velocities of runaway stars. 

\section{Discussion}
\label{section:discussion}

\subsection{Persistence of runaway groups through projection effects}

\begin{figure}[t]
\centering
    \includegraphics[width=0.9\hsize]{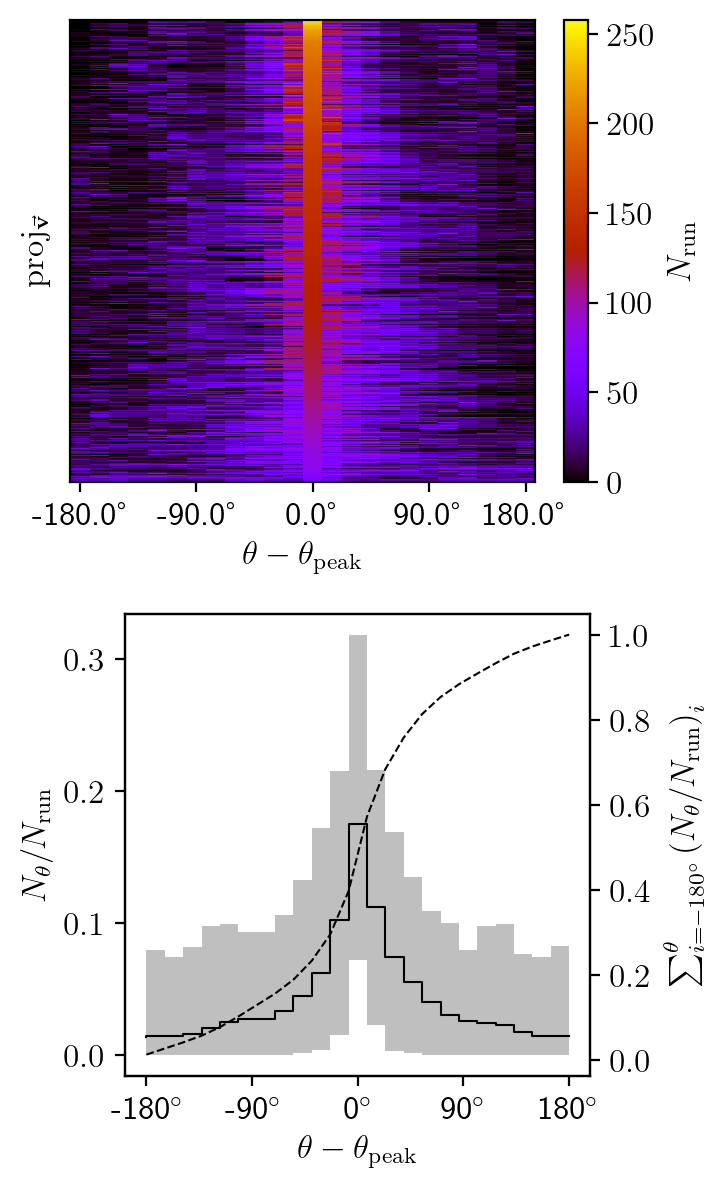}
    \caption{Histograms of runaway star directions from 1,000 different projections. Top: 2D histogram showing the 1,000 projections, sorted vertically and centered horizontally by peak number of runaway stars, $N_{\rm run}$, and wrapped around the x-axis. For every projection there is a strong peak. Looking at each row individually, a fainter peak can be seen to the right or left of the peak value. Bottom: Average value of the projections (black histogram) and the range from maximum to minimum (gray area), with values given by the left axis. The dashed line shows the cumulative sum of the runaway stars per bin, with values given by the right axis. 
    \label{fig:M6_run_proj}}
\end{figure}

In our simulated cluster, the anisotropy of runaway stars is directly linked to the subcluster merger history. To apply this to observations, we must determine whether the anisotropy can be clearly seen from any given observation angle, and the degree to which the groups of runaways appear to be distinct. We tested this by rotating the cluster about a randomly generated unit vector by a randomly generated angle and then recalculating the escape directions. Figure \ref{fig:M6_run_proj} shows a histogram of 1000 random projections (top)  and histograms of the average, maximum, and minimum values of the projections as well as the cumulative sum of runaway fractions per bin  (bottom). The bins are centered such that the central bin is the most prevalent escape angle for each projection. The histograms are strongly peaked, and in almost every projection (each row in the top panel) a fainter secondary peak in the escape direction is still visible. The secondary peak indicates a second SCES runaway group. The strongly peaked histograms, paired with the steep cumulative sum distribution, tells us that if there is a significant subcluster merger history in a cluster resulting in a population of SCES runaways, observers will see an anisotropic distribution of runaway stars regardless of viewing angle.

\subsection{Fossils of cluster assembly history}

Groups of runaways ejected by the SCES are distinctly grouped kinematically, temporally, and directionally. We can identify clusters ejected by this mechanism by finding corresponding grouping in age-velocity space and ${\phi}_{\vec{v}}$-${\theta}_{\vec{v}}$ space. With well-resolved velocities and ages of runaway stars from Gaia, this technique can be applied to find runaways originating from the same subcluster. If paired with other groups of SCES runaways, one can unwind the dynamical assembly history of the natal cluster. Furthermore, in the case of multiple runaway groups, the relative size and velocity of each group can indicate which subcluster merger happened first. The groups with more stars at higher velocities likely were ejected later. Since SCES groups can only be ejected while the central potential is contracting, the kinetic ages of multiple SCES groups can constrain the timescale of their natal cluster's assembly. 

Observations typically only look for runaway O and B-type stars because of their short lifetime. If an O or B star is far from a star forming region, it must be a runaway whereas low-mass stars could just be field stars. Furthermore, there is a bias toward massive stars for ejection via BSS and DES : Massive stars have a multiplicity fraction of $>90\%$ \citep{Sana2014ApJS..215...15S}, and because of mass segregation massive stars are clustered in the stellar core and are more likely to have close dynamical encounters \citep{Oh2016A&A...590A.107O}. Therefore, the presence of many low-mass runaway stars is a unique feature of the SCES. There are 3 O stars ($\geq15\rm~M_\odot$) in group 1 and 7 O stars in group 2, so these SCES runaway groups would be detectable using just the O stars. In the case of anisotropic runaways, we predict these high-mass runaways to be accompanied by many more low-mass stars with similar velocities and ages. Depending on how many massive stars are in a runaway grouping, one can estimate how many low-mass stars accompany them using the IMF. 

\subsection{Observational examples}

\response{The most promising example of the SCES is a group of runaways ejected to the north of R136 in the 30 Doradus star-forming region \citep{Stoop_submitted}. Of the 18 runaways ejected $<1\rm~Myr$ ago, 16 are ejected in the same direction. Additionally, there is an older group of runaways ejected more isotropically, which suggests that they were ejected by DES or BSS. There is a distinct separation in age-velocity space between the SCES and BSS/DES groups, just as we see in our model. This also confirms detectability of SCES runaways using age and velocity. \citet{Stoop_submitted} suggest that the anisotropic runaways were ejected by an interaction with another cluster. Our results further suggest that these ejected stars are the tidal tails produced in the tidal disruption of a late-forming, infalling subcluster. Observations show that there is an ongoing merger between two subclusters within R136 \citep{Sabbi2012ApJ...754L..37S}. This confirms that R136 formed via hierarchical assembly and supports the case for the SCES as the ejection mechanism of its northern runaway star group.}

\response{Another example of possible SCES ejection is seen in the runaway} OB stars ejected from the young massive cluster NGC 6618\response{, which} seem to have a preferential direction. In Fig. 6 of \citet{Stoop2024A&A...681A..21S}, the directions of runaways are plotted showing 7 of 13 stars ejected in a $<90^\circ$ region of the sky. Furthermore, in Fig. 8 of \citet{Stoop2024A&A...681A..21S}, there is a distinct grouping of stars in velocity and stellar age. These stars are ejected from a cluster $\leq1\rm\, Myr$ old \citep{Hanson1997ApJ...489..698H,Hoffmeister2008ApJ...686..310H,Povich2009ApJ...696.1278P,RamirezTannus2017A&A...604A..78R}, which indicates a dynamical ejection, as no stars have exploded as SNe yet. We argue that this group of runaway stars was ejected via subcluster merger rather than individual dynamical interactions.

Figure 12 of \citet{Drew2021MNRAS.508.4952D} shows the ejection directions of runaway OB stars in NGC 360 and Westerlund 2, colored according to the time of their ejection. The runaway populations from both clusters display some anisotropy, with clusters of runaways moving in roughly the same direction. While it is possible these have all been ejected by the DES, we argue the anisotropy indicates a subcluster merger as their origin. 

\subsection{Runaway binary stars}

\begin{figure}[t]
\centering
    \includegraphics[width=1\hsize]{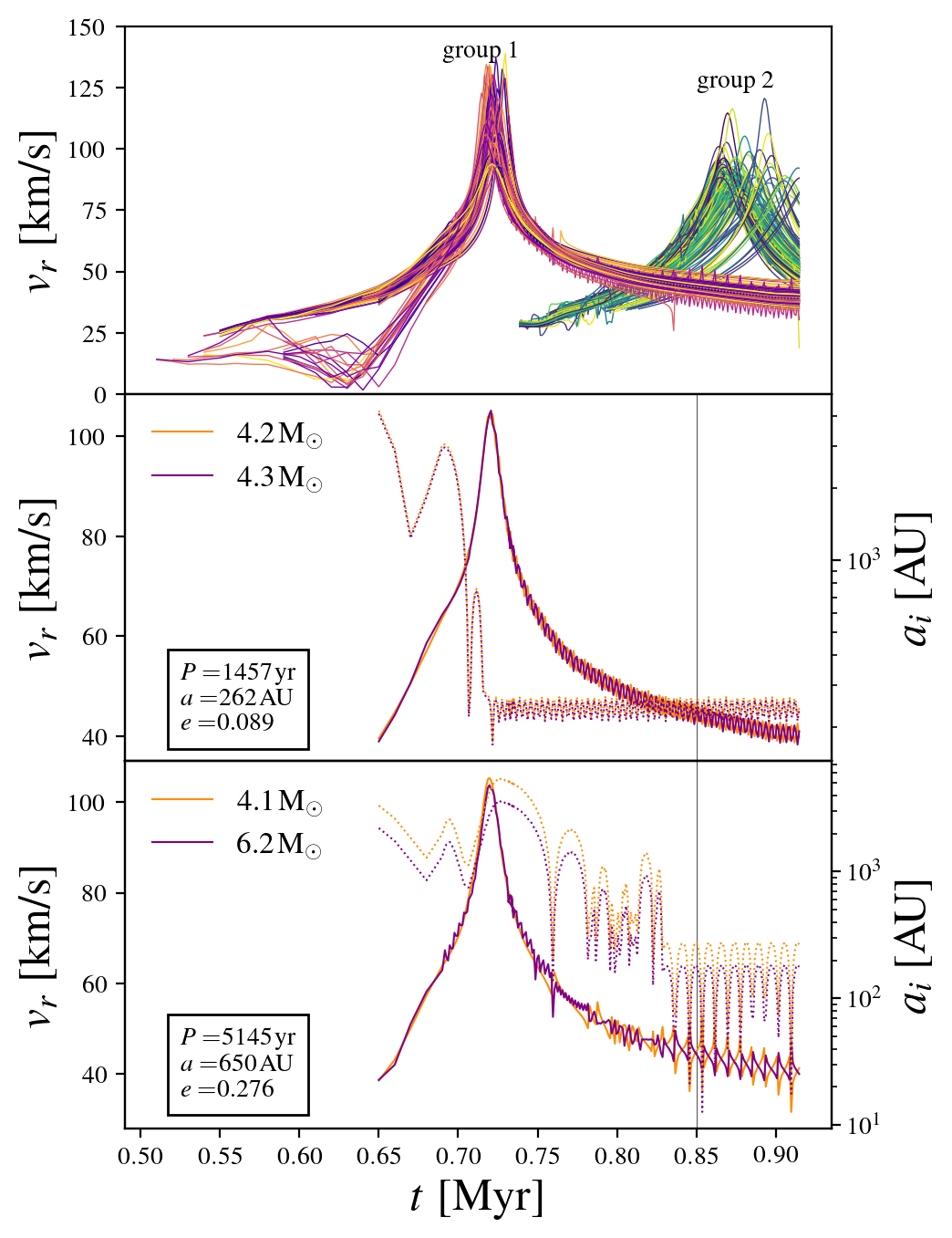}
    \caption{Radial velocity, $v_r,$ relative to the center of mass over time and the orbital properties of the SCES runaway singles and binaries. Top: Time evolution of all the runaways in groups 1 and 2. Middle and bottom: Velocity (solid) and semimajor axes, $a_i$ (dotted) of each member in the two runaway binary stars formed in our simulation. Both are from group 1. The dark lines are the primary stars, and the light lines are the companion stars. Orbital properties for each system are listed in the corresponding panel, calculated for the stable orbits after $t=0.85\rm\, Myr$, which is indicated by the vertical black line.}
    \label{fig:binaries}
\end{figure}

Two dynamical binaries\footnote{\response{We refer to binaries as dynamical if they formed dynamically rather than primordially---in the protostellar disk. All binaries in our model are dynamical.}} were ejected via the SCES in our model, both in group 1. We note that our model did not include any primordial binaries, a topic we will explore in future work. The time evolution of their velocities and orbital separations from the system center of mass is shown in Fig.~\ref{fig:binaries} along with the velocity of all SCES runaways. Both systems formed as extremely wide dynamical binaries. After passing through the central potential, they became significantly harder as \response{orbital energy is lost during the dynamical encounter(s) in the dense core.} This suggests that tight primordial binaries ejected via the SCES are unlikely to be ionized. On the contrary, we show that ejection through the SCES can further harden binaries. This needs to be confirmed by analyzing a similar model with primordial binaries.

The hardening effect occurred in both a circular and a slightly eccentric binary. One binary \response{transitioned into a circular ($e=0.089$) orbit after passing through the central potential, gaining orbital angular momentum through dynamical encounters in the dense core.} The other was irregular, settling into a steady eccentric ($e=0.276$) orbit roughly $\approx0.1\rm\, Myr$ after peak acceleration. 

With the expectation that binaries are preserved and hardened through this mechanism, the SCES has the potential to produce runaway binaries with a broad range of properties. This is because runaways produced by the SCES are a mostly uniform sampling of the stars formed in a subcluster. 

Observations of runaway binaries are fairly rare. \response{The binary fraction of observed OB runaway stars is $\approx8\%$ \citep{Gies1986ApJS...61..419G,Mason1998AJ....115..821M}. Simulations of equal mass binary-binary collisions also found that $\simeq10\%$ of O type runaways were binaries \citep{Leonard1990AJ.....99..608L}. The ejection mechanism producing the observed runaway binaries is not always certain.}

The ejection mechanism is unknown for two confirmed runaway O-star binaries HD 14633 and HD 15137\footnote{HD 14633 and HD 15137 originated in the open cluster NGC 654, approximately 2,400~pc away.}\citep{Gies1986ApJS...61..419G}. They both are in a tight ($P\approx15,30$), eccentric ($e\approx0.7,0.48$) orbits with low mass companions (1-3~M$_\odot$) \citep{McSwain2007ApJ...660..740M}. The low-mass companions could be neutron stars, but neither pulsars nor X-ray emissions are detected, so ejection via BSS cannot be demonstrated\footnote{\citet{McSwain2007ApJ...660..740M} explored the possibility of quiescent neutron stars as the companions. Quiescent neutron stars emit X-rays at a comparable magnitude to O stars but are spectrally distinguishable. X-ray observations with spectral resolution must be done to determine the presence of a quiescent neutron star companion.}. This leaves the case of a low-mass star companion. Most primordial O star binaries with low-mass companions\footnote{The low-mass companions could also have been massive stars stripped of their envelopes via mass-transfer \citep{Sana2012Sci...337..444S}.} have wide orbits ($a>100~\rm AU$) \citep{Sana2014ApJS..215...15S,Moe2017ApJS..230...15M}.  Given our results, it is possible that HD 14633 and HD 15137 originated as wide O-B pairs that were hardened and ejected via the SCES. 

The SCES ejects binary systems while they are young, and they most likely survive. This means binaries ejected via the SCES can potentially undergo two velocity kicks, with a second kick from the BSS following the SCES\footnote{\response{A two step ejection method via DES-BSS was introduced first by \citet{Pflamm2010MNRAS.404.1564P}.}}. Furthermore, because the binaries are hardened during the SCES, the velocity kick from the BSS will be even higher. This could be a channel for producing hypervelocity stars (HVSs). HVSs are stars unbound from the Galaxy, which requires $v\gtrsim400\rm~km~s^{-1}$ in the Galactic rest frame depending on location and direction. \citet{Tauris2015MNRAS.448L...6T} found the BSS capable of producing HVSs. For HVSs with masses $M_\star=0.9,~3.5,{\rm\ and\ } 10\rm~M_\odot$ they found kick velocities in the Galactic rest frame up to $v^{\rm max}_{grf}=1,280,~770,{\rm\ and\ }550~\rm km~s^{-1}$, respectively. These maximum velocities are only possible via the BSS for particularly favorable binary parameters, such as closer orbits. 

Binaries ejected via the SCES become more favorable HVS progenitors via subsequent BSS: they are closer in orbit, further from the galactic disk, and already locally unbound. Therefore, SCES binaries are more likely to produce HVSs via a subsequent BSS than if a binary undergoes the BSS while still bound to its parent cluster. This two-step mechanism is a likely channel for producing HVSs with trajectories not pointing toward the supermassive black hole at the Galactic center\footnote{HVSs can be produced when a tight binary has a close encounter with the supermassive black hole at the center of the Galaxy and its companion is captured. A large number of late B-type HVSs originate in this way \citep{Brown2014ApJ...787...89B}.}\citep{Hills1988Natur.331..687H}. 

\response{The runaway binaries produced in our model could not become HVSs, as their orbital velocities are too low. Rather, we argue that a close primordial binary could become further hardened by ejection through the SCES and thus acquire a fast enough orbital velocity to become a HVS after the SN of its companion.}


\subsection{Importance of initial conditions}

Our initial conditions are simplified: a spherical cloud of almost uniform density with isotropic turbulence. In reality, GMCs forming star clusters are filamentary \citep[see][]{Heyer2015ARA&A..53..583H,Klessen2016SAAS...43...85K,Hacar2023}. Subclusters forming from spherical clouds are closer to the center of mass and more evenly distributed than those forming from filamentary clouds. Accounting for filamentary initial conditions could dramatically increase the effectiveness of the SCES.

The two subclusters that formed furthest from the center of the cluster in M6 were ejected. Subclusters that form late or far enough away that they approach the assembling cluster after most of the other subclusters have already merged will be most effectively accelerated in the SCES. The subcluster must approach during the period when the central potential is still contracting. With more realistic filamentary initial conditions, there will be more subclusters formed as the dense gas will be more distributed. We therefore predict a much higher fraction of SCES runaways from star cluster models with more realistic initial conditions due to the increase in the number and infall distance of subcluster merger events.

In future work, we plan to import GMCs formed in large-scale galaxy simulations into \textsc{torch} to assess the extent to which realistic initial GMCs affect the fraction of runaway stars formed via the SCES. This will give us a better idea of how many observed runaways can be attributed to the SCES versus the BSS or the DES. This issue highlights the necessity of using realistic initial conditions, particularly for realistic subcluster dynamics in star cluster formation models. 

\response{
\subsection{High peak stellar density}
\label{subsec:peak}
\begin{figure}[t]
\centering
    \includegraphics[width=1\hsize]{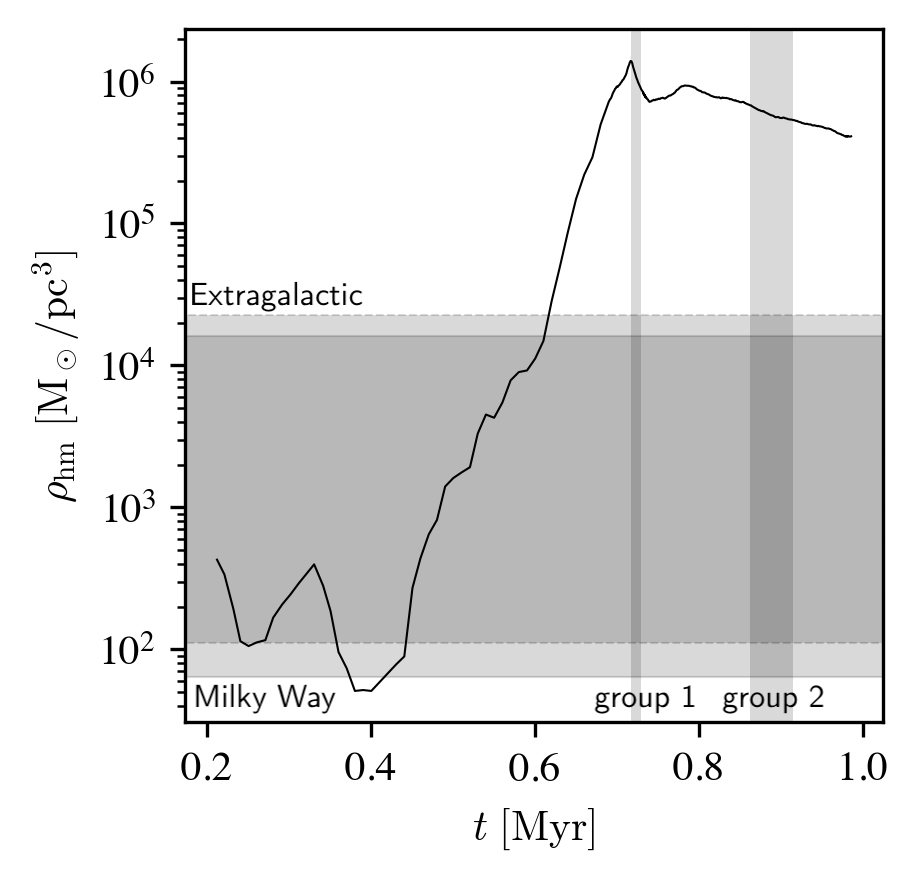}
    \caption{\response{Half-mass stellar density of our model over time. Regions of observed $\rho_{\rm hm}$ values in young clusters ($<10$~Myr) are indicated by horizontal gray shaded regions. The dashed border lines show extragalactic clusters, and the solid border lines show Milky Way clusters. The vertical shaded regions show the time windows when runaways are dynamically ejected at peak velocity in our M6 model. The observational data were taken from \citet{PortegiesZwart2010ARA&A..48..431P}. }}
    \label{fig:rho_hm}
\end{figure}

The stellar encounter rate per unit volume $\gamma$ in a star cluster is given by $\gamma\propto\rho^2/\sigma$ where $\rho$ is the stellar density and $\sigma$ is the velocity dispersion \citep{Verbunt1987IAUS..125..187V}. 
Due to the high initial density and low virial parameter of the initial cloud for our M6 model, the star cluster collapses to a high stellar density, resulting in a high stellar encounter rate. We compare the half-mass density of our model to young ($<10\rm~Myr$) Milky Way and Local Group  star clusters in Fig.~\ref{fig:rho_hm}. We also indicate the time periods of ejection (peak velocity) for the two runaway groups.

The stellar density of our model when the runaway groups are ejected is higher than the observed values. However, we note that this is the density just after core collapse. The cluster will relax to a lower stellar density as it continues to evolve, and, indeed, the stellar density has already begun to decrease at a steady rate following the ejection of group 1.  Considering their ages (the youngest is 3~Myr old), the observed clusters used to make Fig. 11 are most likely already further post-collapse than our model. Also, aside from R136, none are as massive as our M6 model, which is more characteristic of clusters in starburst galaxies.

Regardless, the high stellar density in our model compared to observations implies that the interaction rate could be inflated. The fraction of stars ejected via the SCES in clusters that do not reach as high stellar densities during core collapse would likely be lower than in our model. Additional models with more self-consistent initial conditions are needed to constrain the efficiency and frequency of SCES runaways in clusters of varying mass and density.
}

\section{Conclusions}
\label{section:conclusions}

Using a star-by-star simulation of star cluster formation from a gas cloud, we have identified the SCES as a new channel for the origin of runaway stars. This scenario occurs when a subcluster forms late, after the rest of the cluster is mostly assembled, and the subcluster then falls into the contracting central potential, \response{becoming tidally disrupted and} ejecting the majority of the stars in the subcluster as runaways \response{in tidal tails}. 

We believe the reason this phenomenon had not been identified earlier is the oversimplification of initial conditions in many star cluster formation simulations. Using a smooth spherical cloud instead of a cloud more realistically structured by supersonic turbulence results in less energetic subcluster mergers since star formation is more centrally concentrated in a spherical cloud. This emphasizes the need for linking scales and using realistic clouds formed in galaxy formation simulations as initial conditions for models of cluster formation. 

To determine the fraction of runaways formed by the SCES, BSS, and DES, we must import a self-consistently formed GMC to model the hierarchical assembly correctly. We also need to include primordial binaries, as they are essential for ejection via the DES or BSS. Furthermore, the simulations must be run for long enough to account for long-term dynamics and the timing of SNe. The M6 simulation in this study was only run for $1.35\, t_{\rm ff}$, equivalent to less than $1~$Myr, and therefore no SNe occurred during this time.

The runaways formed by the SCES are highly correlated with respect to their velocities, ages, and ejection directions. Finding groupings of runaway stars with similar values of these three properties is a straightforward method for detecting runaways formed from the same subcluster merger. If anisotropy is observed in a group of runaway stars, this indicates a distinct subcluster merger history. Conversely, if the runaways are isotropic, this indicates a mild cluster assembly history and ejection primarily via the DES or BSS channels. A multimodal spatial or velocity distribution of runaway stars could indicate multiple distinct subcluster mergers, as seen in our model. Depending on the resolution of observations, these groups of runaways can be used as fossils to trace the assembly history of a star cluster. 

An important caveat for confirming an anisotropic distribution of runaway stars concerns the number of detectable runaways: if a cluster ejects a small number of runaway stars isotropically, even fewer will be massive and detectable. This could falsely imply anisotropy in runaway star ejection directions. A sufficiently large sample of massive runaway stars is needed to exclude this possibility and confirm the anisotropy of the ejection direction. On the other hand, if grouped runaways in the small sampling have the same velocities and kinetic ages, it is likely that they were ejected by the SCES. 

We surveyed observational work and found several anisotropic populations of runaway stars. \response{\cite{Stoop_submitted} show that the group of runaways moving north of R136 has properties consistent with ejection via the SCES. We also find runaway groups in other clusters that could be produced by the SCES,} though more analysis needs to be done to confirm this. Regardless, observations of runaway stars must be looked at through an additional lens of possible \response{SCES} origins. 

The SCES is capable of producing runaway binaries. Two wide binaries in our model were ejected, and through ejection, their orbits hardened significantly. This suggests that primordial binaries can survive ejection and end up with harder orbits. The SCES can thus explain observations of runaway binaries. This scenario also allows for a subsequent BSS ejection, which will be more energetic than a birthplace BSS as the binary progenitor is already unbound and in a hardened orbit. A two-step SCES-BSS ejection is therefore a potential production mechanism for HVSs unbound from the Galaxy.

\begin{acknowledgements}

\response{We thank the referee Mark Gieles for his insightful comments, one of which led to Sect.~\ref{subsec:peak}.} We thank Yuri Levin for pointing out runaway binary stars as a potential signature of SCES. B.P. was partly supported by a fellowship from the International Max Planck Research School for Astronomy and Cosmic Physics at the University of Heidelberg (IMPRS-HD). M.-M.M.L., B.P., and E.P.A were partly supported by NSF grants AST18-15461 and AST23-07950. \response{E.P.A. also received support from NASA Astrophysical Theory grant 80NSSC24K0935.} C.C.-C. is supported by a Canada Graduate Scholarship - Doctoral (CGS D) from the Natural Sciences and Engineering Research Council of Canada (NSERC). This work used Stampede 2 at TACC through allocation PHY220160 from the Advanced Cyberinfrastructure Coordination Ecosystem: Services \& Support (ACCESS) program, which is supported by National Science Foundation grants 21-38259, 21-38286, 21-38307, 21-37603, and 21-38296. The code development that facilitated this study was done on Snellius through the Dutch National Supercomputing Center SURF grants 15220 and 2023/ENW/01498863. S.M.A. acknowledges the support of NSF grant AST-2009679. R.S.K.\ and S.C.O.G.\ acknowledge financial support from the European Research Council via the ERC Synergy Grant ``ECOGAL'' (project ID 855130),  from the Heidelberg Cluster of Excellence (EXC 2181 - 390900948) ``STRUCTURES'', funded by the German Excellence Strategy, and from the German Ministry for Economic Affairs and Climate Action in project ``MAINN'' (funding ID 50OO2206). The team in Heidelberg also thanks The L\"{a}nd and the German Science Foundation (DFG) for computing resources provided in bwHPC supported by grant INST 35/1134-1 FUGG and for data storage at SDS@hd supported by grant INST 35/1314-1 FUGG. The authors acknowledge Interstellar Institute's program ``II6'' and the Paris-Saclay University's Institut Pascal for hosting discussions that nourished the development of the ideas behind this work.

\end{acknowledgements}

\bibliographystyle{aa}
\bibliography{references}

\begin{thebibliography}{59}
\expandafter\ifx\csname natexlab\endcsname\relax\def\natexlab#1{#1}\fi

\bibitem[{Baczynski {et~al.}(2015)Baczynski, Glover, \& Klessen}]{FERVENT10.1093/mnras/stv1906}
Baczynski, C., Glover, S. C.~O., \& Klessen, R.~S. 2015, Monthly Notices of the Royal Astronomical Society, 454, 380

\bibitem[{{Bate} {et~al.}(1995){Bate}, {Bonnell}, \& {Price}}]{Bate1995MNRAS.277..362B}
{Bate}, M.~R., {Bonnell}, I.~A., \& {Price}, N.~M. 1995, \mnras, 277, 362

\bibitem[{{Blaauw}(1961)}]{Blaauw1961BAN....15..265B}
{Blaauw}, A. 1961, \bain, 15, 265

\bibitem[{{Blaauw} \& {Morgan}(1954)}]{Blaauw1954ApJ...119..625B}
{Blaauw}, A. \& {Morgan}, W.~W. 1954, \apj, 119, 625

\bibitem[{{Brown} {et~al.}(2014){Brown}, {Geller}, \& {Kenyon}}]{Brown2014ApJ...787...89B}
{Brown}, W.~R., {Geller}, M.~J., \& {Kenyon}, S.~J. 2014, \apj, 787, 89

\bibitem[{{Colella} \& {Woodward}(1984)}]{Colella1984JCoPh..54..174C}
{Colella}, P. \& {Woodward}, P.~R. 1984, Journal of Computational Physics, 54, 174

\bibitem[{{Cournoyer-Cloutier} {et~al.}(2023){Cournoyer-Cloutier}, {Sills}, {Harris}, {Appel}, {Lewis}, {Polak}, {Tran}, {Wilhelm}, {Mac Low}, {McMillan}, \& {Portegies Zwart}}]{Cournoyer-Cloutier2023MNRAS.521.1338C}
{Cournoyer-Cloutier}, C., {Sills}, A., {Harris}, W.~E., {et~al.} 2023, \mnras, 521, 1338

\bibitem[{{Cournoyer-Cloutier} {et~al.}(2021){Cournoyer-Cloutier}, {Tran}, {Lewis}, {Wall}, {Harris}, {Mac Low}, {McMillan}, {Portegies Zwart}, \& {Sills}}]{2021Cournoyer-Cloutier}
{Cournoyer-Cloutier}, C., {Tran}, A., {Lewis}, S., {et~al.} 2021, MNRAS, 501, 4464

\bibitem[{{Drew} {et~al.}(2021){Drew}, {Mongui{\'o}}, \& {Wright}}]{Drew2021MNRAS.508.4952D}
{Drew}, J.~E., {Mongui{\'o}}, M., \& {Wright}, N.~J. 2021, \mnras, 508, 4952

\bibitem[{Federrath {et~al.}(2010)Federrath, Banerjee, Clark, \& Klessen}]{Federrath_2010}
Federrath, C., Banerjee, R., Clark, P.~C., \& Klessen, R.~S. 2010, The Astrophysical Journal, 713, 269

\bibitem[{{Fryxell} {et~al.}(2000){Fryxell}, {Olson}, {Ricker}, {Timmes}, {Zingale}, {Lamb}, {MacNeice}, {Rosner}, {Truran}, \& {Tufo}}]{flash}
{Fryxell}, B., {Olson}, K., {Ricker}, P., {et~al.} 2000, ApJs, 131, 273

\bibitem[{{Fujii} {et~al.}(2007){Fujii}, {Iwasawa}, {Funato}, \& {Makino}}]{Fujii2007PASJ...59.1095F}
{Fujii}, M., {Iwasawa}, M., {Funato}, Y., \& {Makino}, J. 2007, \pasj, 59, 1095

\bibitem[{{Fujii} \& {Portegies Zwart}(2011)}]{Fujii2011Sci...334.1380F}
{Fujii}, M.~S. \& {Portegies Zwart}, S. 2011, Science, 334, 1380

\bibitem[{{Gies} \& {Bolton}(1986)}]{Gies1986ApJS...61..419G}
{Gies}, D.~R. \& {Bolton}, C.~T. 1986, \apjs, 61, 419

\bibitem[{{Goodwin} {et~al.}(2004){Goodwin}, {Whitworth}, \& {Ward-Thompson}}]{Goodwin2004A&A...414..633G}
{Goodwin}, S.~P., {Whitworth}, A.~P., \& {Ward-Thompson}, D. 2004, \aap, 414, 633

\bibitem[{{Grudi{\'c}} {et~al.}(2018){Grudi{\'c}}, {Guszejnov}, {Hopkins}, {Lamberts}, {Boylan-Kolchin}, {Murray}, \& {Schmitz}}]{Grudic2018MNRAS.481..688G}
{Grudi{\'c}}, M.~Y., {Guszejnov}, D., {Hopkins}, P.~F., {et~al.} 2018, \mnras, 481, 688

\bibitem[{{Hacar} {et~al.}(2023){Hacar}, {Clark}, {Heitsch}, {Kainulainen}, {Panopoulou}, {Seifried}, \& {Smith}}]{Hacar2023}
{Hacar}, A., {Clark}, S.~E., {Heitsch}, F., {et~al.} 2023, in Astronomical Society of the Pacific Conference Series, Vol. 534, Protostars and Planets VII, ed. S.~{Inutsuka}, Y.~{Aikawa}, T.~{Muto}, K.~{Tomida}, \& M.~{Tamura}, 153

\bibitem[{{Hanson} {et~al.}(1997){Hanson}, {Howarth}, \& {Conti}}]{Hanson1997ApJ...489..698H}
{Hanson}, M.~M., {Howarth}, I.~D., \& {Conti}, P.~S. 1997, \apj, 489, 698

\bibitem[{{Heyer} \& {Dame}(2015)}]{Heyer2015ARA&A..53..583H}
{Heyer}, M. \& {Dame}, T.~M. 2015, \araa, 53, 583

\bibitem[{{Hills}(1988)}]{Hills1988Natur.331..687H}
{Hills}, J.~G. 1988, \nat, 331, 687

\bibitem[{{Hoffmeister} {et~al.}(2008){Hoffmeister}, {Chini}, {Scheyda}, {Schulze}, {Watermann}, {N{\"u}rnberger}, \& {Vogt}}]{Hoffmeister2008ApJ...686..310H}
{Hoffmeister}, V.~H., {Chini}, R., {Scheyda}, C.~M., {et~al.} 2008, \apj, 686, 310

\bibitem[{{Hoogerwerf} {et~al.}(2000){Hoogerwerf}, {de Bruijne}, \& {de Zeeuw}}]{Hoogerwerf2000ApJ...544L.133H}
{Hoogerwerf}, R., {de Bruijne}, J.~H.~J., \& {de Zeeuw}, P.~T. 2000, \apjl, 544, L133

\bibitem[{{Klessen} \& {Glover}(2016)}]{Klessen2016SAAS...43...85K}
{Klessen}, R.~S. \& {Glover}, S. C.~O. 2016, in Saas-Fee Advanced Course, Vol.~43, Saas-Fee Advanced Course, ed. Y.~{Revaz}, P.~{Jablonka}, R.~{Teyssier}, \& L.~{Mayer}, 85

\bibitem[{{Kolmogorov}(1941)}]{Kolmogorov1941DoSSR..30..301K}
{Kolmogorov}, A. 1941, Akademiia Nauk SSSR Doklady, 30, 301

\bibitem[{{Kroupa}(2002)}]{2002Sci...295...82Kroupa}
{Kroupa}, P. 2002, Science, 295, 82

\bibitem[{{Leonard} \& {Duncan}(1990)}]{Leonard1990AJ.....99..608L}
{Leonard}, P. J.~T. \& {Duncan}, M.~J. 1990, \aj, 99, 608

\bibitem[{{Lewis} {et~al.}(2023){Lewis}, {McMillan}, {Low}, {Cournoyer-Cloutier}, {Polak}, {Wilhelm}, {Tran}, {Sills}, {Portegies Zwart}, {Klessen}, \& {Wall}}]{Lewis2023ApJ...944..211L}
{Lewis}, S.~C., {McMillan}, S. L.~W., {Low}, M.-M.~M., {et~al.} 2023, \apj, 944, 211

\bibitem[{{Lucas} {et~al.}(2018){Lucas}, {Rybak}, {Bonnell}, \& {Gieles}}]{Lucas2018MNRAS.474.3582L}
{Lucas}, W.~E., {Rybak}, M., {Bonnell}, I.~A., \& {Gieles}, M. 2018, \mnras, 474, 3582

\bibitem[{{Mason} {et~al.}(1998){Mason}, {Gies}, {Hartkopf}, {Bagnuolo}, {ten Brummelaar}, \& {McAlister}}]{Mason1998AJ....115..821M}
{Mason}, B.~D., {Gies}, D.~R., {Hartkopf}, W.~I., {et~al.} 1998, \aj, 115, 821

\bibitem[{{McSwain} {et~al.}(2007){McSwain}, {Ransom}, {Boyajian}, {Grundstrom}, \& {Roberts}}]{McSwain2007ApJ...660..740M}
{McSwain}, M.~V., {Ransom}, S.~M., {Boyajian}, T.~S., {Grundstrom}, E.~D., \& {Roberts}, M. S.~E. 2007, \apj, 660, 740

\bibitem[{{Miyoshi} \& {Kusano}(2005)}]{Miyoshi2005JCoPh.208..315M}
{Miyoshi}, T. \& {Kusano}, K. 2005, Journal of Computational Physics, 208, 315

\bibitem[{{Moe} \& {Di Stefano}(2017)}]{Moe2017ApJS..230...15M}
{Moe}, M. \& {Di Stefano}, R. 2017, \apjs, 230, 15

\bibitem[{{Oh} \& {Kroupa}(2016)}]{Oh2016A&A...590A.107O}
{Oh}, S. \& {Kroupa}, P. 2016, \aap, 590, A107

\bibitem[{{Pelupessy} {et~al.}(2013){Pelupessy}, {van Elteren}, {de Vries}, {McMillan}, {Drost}, \& {Portegies Zwart}}]{amuse}
{Pelupessy}, F.~I., {van Elteren}, A., {de Vries}, N., {et~al.} 2013, A\&A, 557, A84

\bibitem[{{Pflamm-Altenburg} \& {Kroupa}(2010)}]{Pflamm2010MNRAS.404.1564P}
{Pflamm-Altenburg}, J. \& {Kroupa}, P. 2010, \mnras, 404, 1564

\bibitem[{{Polak} {et~al.}(2023){Polak}, {Mac Low}, {Klessen}, {Teh}, {Cournoyer-Cloutier}, {Andersson}, {Appel}, {Tran}, {Lewis}, {Wilhelm}, {Portegies Zwart}, {Glover}, {Wang}, \& {McMillan}}]{Polak2023}
{Polak}, B., {Mac Low}, M.-M., {Klessen}, R.~S., {et~al.} 2023, arXiv e-prints, arXiv:2312.06509

\bibitem[{Portegies~Zwart \& McMillan(2018)}]{amusebook}
Portegies~Zwart, S. \& McMillan, S. 2018, Astrophysical Recipes, 2514-3433 (IOP Publishing)

\bibitem[{{Portegies Zwart} {et~al.}(2009){Portegies Zwart}, McMillan, Harfst, Groen, Fujii, NuallÃ¡in, Glebbeek, Heggie, Lombardi, Hut, Angelou, Banerjee, Belkus, Fragos, Fregeau, Gaburov, Izzard, JuriÄ, Justham, Sottoriva, Teuben, {van Bever}, Yaron, \& Zemp}]{PORTEGIESZWART2009369amuse1}
{Portegies Zwart}, S., McMillan, S., Harfst, S., {et~al.} 2009, New Astronomy, 14, 369

\bibitem[{{Portegies Zwart} {et~al.}(2013){Portegies Zwart}, {McMillan}, {van Elteren}, {Pelupessy}, \& {de Vries}}]{Portegies2013CoPhC.184..456Pamuse2}
{Portegies Zwart}, S., {McMillan}, S.~L.~W., {van Elteren}, E., {Pelupessy}, I., \& {de Vries}, N. 2013, Computer Physics Communications, 184, 456

\bibitem[{{Portegies Zwart} {et~al.}(2010){Portegies Zwart}, {McMillan}, \& {Gieles}}]{PortegiesZwart2010ARA&A..48..431P}
{Portegies Zwart}, S.~F., {McMillan}, S. L.~W., \& {Gieles}, M. 2010, \araa, 48, 431

\bibitem[{{Portegies Zwart} \& {Verbunt}(1996)}]{seba}
{Portegies Zwart}, S.~F. \& {Verbunt}, F. 1996, \aap, 309, 179

\bibitem[{{Poveda} {et~al.}(1967){Poveda}, {Ruiz}, \& {Allen}}]{Poveda1967BOTT....4...86P}
{Poveda}, A., {Ruiz}, J., \& {Allen}, C. 1967, Boletin de los Observatorios Tonantzintla y Tacubaya, 4, 86

\bibitem[{{Povich} {et~al.}(2009){Povich}, {Churchwell}, {Bieging}, {Kang}, {Whitney}, {Brogan}, {Kulesa}, {Cohen}, {Babler}, {Indebetouw}, {Meade}, \& {Robitaille}}]{Povich2009ApJ...696.1278P}
{Povich}, M.~S., {Churchwell}, E., {Bieging}, J.~H., {et~al.} 2009, \apj, 696, 1278

\bibitem[{{Rahner} {et~al.}(2017){Rahner}, {Pellegrini}, {Glover}, \& {Klessen}}]{Rahner2017MNRAS.470.4453R}
{Rahner}, D., {Pellegrini}, E.~W., {Glover}, S. C.~O., \& {Klessen}, R.~S. 2017, \mnras, 470, 4453

\bibitem[{{Rahner} {et~al.}(2019){Rahner}, {Pellegrini}, {Glover}, \& {Klessen}}]{Rahner2019MNRAS.483.2547R}
{Rahner}, D., {Pellegrini}, E.~W., {Glover}, S. C.~O., \& {Klessen}, R.~S. 2019, \mnras, 483, 2547

\bibitem[{{Ram{\'\i}rez-Tannus} {et~al.}(2017){Ram{\'\i}rez-Tannus}, {Kaper}, {de Koter}, {Tramper}, {Bik}, {Ellerbroek}, {Ochsendorf}, {Ram{\'\i}rez-Agudelo}, \& {Sana}}]{RamirezTannus2017A&A...604A..78R}
{Ram{\'\i}rez-Tannus}, M.~C., {Kaper}, L., {de Koter}, A., {et~al.} 2017, \aap, 604, A78

\bibitem[{{Ricker}(2008)}]{Ricker2008ApJS..176..293R}
{Ricker}, P.~M. 2008, \apjs, 176, 293

\bibitem[{{Sabbi} {et~al.}(2012){Sabbi}, {Lennon}, {Gieles}, {de Mink}, {Walborn}, {Anderson}, {Bellini}, {Panagia}, {van der Marel}, \& {Ma{\'\i}z Apell{\'a}niz}}]{Sabbi2012ApJ...754L..37S}
{Sabbi}, E., {Lennon}, D.~J., {Gieles}, M., {et~al.} 2012, \apjl, 754, L37

\bibitem[{{Sana} {et~al.}(2012){Sana}, {de Mink}, {de Koter}, {Langer}, {Evans}, {Gieles}, {Gosset}, {Izzard}, {Le Bouquin}, \& {Schneider}}]{Sana2012Sci...337..444S}
{Sana}, H., {de Mink}, S.~E., {de Koter}, A., {et~al.} 2012, Science, 337, 444

\bibitem[{{Sana} {et~al.}(2014){Sana}, {Le Bouquin}, {Lacour}, {Berger}, {Duvert}, {Gauchet}, {Norris}, {Olofsson}, {Pickel}, {Zins}, {Absil}, {de Koter}, {Kratter}, {Schnurr}, \& {Zinnecker}}]{Sana2014ApJS..215...15S}
{Sana}, H., {Le Bouquin}, J.~B., {Lacour}, S., {et~al.} 2014, \apjs, 215, 15

\bibitem[{{Sana} {et~al.}(2022){Sana}, {Ram{\'\i}rez-Agudelo}, {H{\'e}nault-Brunet}, {Mahy}, {Almeida}, {de Koter}, {Bestenlehner}, {Evans}, {Langer}, {Schneider}, {Crowther}, {de Mink}, {Herrero}, {Lennon}, {Gieles}, {Ma{\'\i}z Apell{\'a}niz}, {Renzo}, {Sabbi}, {van Loon}, \& {Vink}}]{Sana2022A&A...668L...5S}
{Sana}, H., {Ram{\'\i}rez-Agudelo}, O.~H., {H{\'e}nault-Brunet}, V., {et~al.} 2022, \aap, 668, L5

\bibitem[{{Stoop} {et~al.}(2024{\natexlab{a}}){Stoop}, {de Koter}, {Kaper}, {Brands}, {Zwart}, {Sana}, {Stoppa}, {Gieles}, {Mahy}, {Shenar}, {Guo}, {Nelemans}, \& {Rieder}}]{Stoop_submitted}
{Stoop}, M., {de Koter}, A., {Kaper}, L., {et~al.} 2024{\natexlab{a}}, \nat, subm.

\bibitem[{{Stoop} {et~al.}(2024{\natexlab{b}}){Stoop}, {Derkink}, {Kaper}, {de Koter}, {Rogers}, {Ram{\'\i}rez-Tannus}, {Guo}, \& {Azatyan}}]{Stoop2024A&A...681A..21S}
{Stoop}, M., {Derkink}, A., {Kaper}, L., {et~al.} 2024{\natexlab{b}}, \aap, 681, A21

\bibitem[{{Tauris}(2015)}]{Tauris2015MNRAS.448L...6T}
{Tauris}, T.~M. 2015, \mnras, 448, L6

\bibitem[{{Verbunt} \& {Hut}(1987)}]{Verbunt1987IAUS..125..187V}
{Verbunt}, F. \& {Hut}, P. 1987, in The Origin and Evolution of Neutron Stars, ed. D.~J. {Helfand} \& J.~H. {Huang}, Vol. 125, 187

\bibitem[{{Wall} {et~al.}(2020){Wall}, {Mac Low}, {McMillan}, {Klessen}, {Portegies Zwart}, \& {Pellegrino}}]{2020Wall}
{Wall}, J.~E., {Mac Low}, M.-M., {McMillan}, S. L.~W., {et~al.} 2020, ApJ, 904, 192

\bibitem[{{Wall} {et~al.}(2019){Wall}, {McMillan}, {Mac Low}, {Klessen}, \& {Portegies Zwart}}]{wall2019}
{Wall}, J.~E., {McMillan}, S. L.~W., {Mac Low}, M.-M., {Klessen}, R.~S., \& {Portegies Zwart}, S. 2019, ApJ, 887, 62

\bibitem[{{Wang} {et~al.}(2020){Wang}, {Iwasawa}, {Nitadori}, \& {Makino}}]{petar}
{Wang}, L., {Iwasawa}, M., {Nitadori}, K., \& {Makino}, J. 2020, MNRAS, 497, 536

\bibitem[{{Wilhelm} {et~al.}(2023){Wilhelm}, {Portegies Zwart}, {Cournoyer-Cloutier}, {Lewis}, {Polak}, {Tran}, \& {Mac Low}}]{Wilhelm2023MNRAS.520.5331W}
{Wilhelm}, M. J.~C., {Portegies Zwart}, S., {Cournoyer-Cloutier}, C., {et~al.} 2023, \mnras, 520, 5331

\end{thebibliography}

\end{document}